\documentclass[aps,prl,preprint,groupedaddress,showpacs]{revtex4}

\usepackage{graphicx}
\usepackage{dcolumn}
\usepackage{bm}

\begin{document}

\preprint{APS/123-QED}

\title{Field-Driven Domain-Wall Dynamics in GaMnAs Films with Perpendicular Anisotropy}

\author{A. Dourlat$^{a,b}$, V. Jeudy$^{a,c}$, A. Lema\^{\i}tre$^d$, C. Gourdon$^{a,b}$}
 
\affiliation{
$^a$  CNRS, UMR7588, Institut des Nanosciences de Paris, 140 rue de Lourmel, Paris, F-75015 France\\
$^b$ Universit$\acute{e}$ Pierre et Marie Curie-Paris 6, UMR 7588, INSP, Paris, France\\
$^c$ Universit$\acute{e}$ de Cergy-Pontoise, 95000 Cergy-Pontoise, France\\
$^d$ Laboratoire de Photonique et Nanostructures, CNRS, UPR 20\\ Route de Nozay, Marcoussis, F-91460 
France}

\date{\today}

\begin{abstract}

We combine magneto-optical imaging and a magnetic field pulse technique to study domain wall dynamics in a ferromagnetic (Ga,Mn)As layer with perpendicular easy axis. Contrary to ultrathin metallic layers, the depinning field is found to be smaller than the Walker field, thereby allowing for the observation of the steady and precessional flow regimes. The domain wall width and damping parameters are determined self-consistently. The damping, 30 times larger than the one deduced from ferromagnetic resonance, is shown to essentially originate from the non-conservation of the magnetization modulus. An unpredicted damping resonance and a dissipation regime associated with the existence of horizontal Bloch lines are also revealed.
\end{abstract}


\maketitle

In ferromagnetic systems, domain wall (DW) motion driven by a magnetic field~\cite{slonczewski73-74,schryer-walker,hubert-german,malozemoff,leeuw} or a spin-polarized current~\cite{berger,thiavilleEPL2005,yamanouchi-science} presents a variety of dynamical regimes.
Depending on the field strength, several regimes characterized by the dynamics of the magnetization vector inside the DW and the DW mobility (field derivative of the velocity) are predicted to occur. Theoretically, the dissipation-limited regimes were mostly investigated in a system consisting of an ideal ferromagnetic film with uniaxial perpendicular anisotropy, subject to a magnetic field $H$ parallel to the easy axis~\cite{slonczewski73-74,schryer-walker,malozemoff,hubert-german}.
In the Walker steady regime, the DW structure is stationary and a linear velocity $v=\mu H$ is expected up to the Walker field~\cite{schryer-walker,malozemoff}. Above this field the precession of the DW magnetization around $H$ leads to a DW back and forth motion, reducing its average velocity~\cite{hubert-german}. This regime may become unstable, leading to the nucleation and propagation of Bloch lines inside the DW~\cite{slonczewski73-74,thiavilleEPJ}. In the high-field range of the precessional regime the damping torque becomes large enough to move the DW with linear velocity but reduced mobility with respect to the Walker regime.
These various dynamical regimes could be observed in the past in micrometer-thick garnet films~\cite{malozemoff,leeuw} and very recently in nanowires with in-plane magnetization~\cite{beach-nat,hayashi}. However, in ultrathin metallic ferromagnetic films with perpendicular magnetization dissipation-limited regimes are masked by thermally-activated creep and depinning regimes~\cite{lemerle,metaxas}. For instance, in Pt/Co/Pt layers only the high field linear precessional regime could be observed owing to the strong pinning of DWs~\cite{metaxas}.

In ferromagnetic semiconductors, DW dynamics has been explored only recently. The creep regime has been observed in GaMnAs thin films and wires~\cite{tang,yamanouchi-science}. However, the strength of the applied field was too small to reach the dissipation-limited regimes. The observation of these regimes is of prime importance to determine the relevant dissipation processes involved in DW motion. In particular, the nature of the ferromagnetism (hole-mediated interaction between diluted Mn ions~\cite{dietlPRB2001,jungwirth}) could lead to specific features in the DW dynamics. Moreover, the understanding of dissipation processes in GaMnAs should have important implications for the study of current-driven DW motion. It should help to discriminate between the damping contribution and the spin transfer one, which is not yet well understood~\cite{thiavilleEPL2005}. GaMnAs is a good candidate for experimental investigation of this question since the current density required to move a DW was found to be two orders of magnitude smaller than in metallic nanowires~\cite{chiba}.

In this letter, we report on DW dynamics in GaMnAs over a wide range of magnetic field and temperature values. A quantitative analysis of the DW dynamics leads to the identification of the steady and precessional regimes and to a self-consistent determination of the DW width and damping parameters. This damping is shown to involve the variation of the magnetization magnitude. We also identify a dynamical regime consistent with the existence of horizontal Bloch lines. Finally, DW dynamics reveals an unpredicted velocity peak.

\begin{figure}[htbp]
	\begin{center}
	\includegraphics[width=0.98\linewidth]{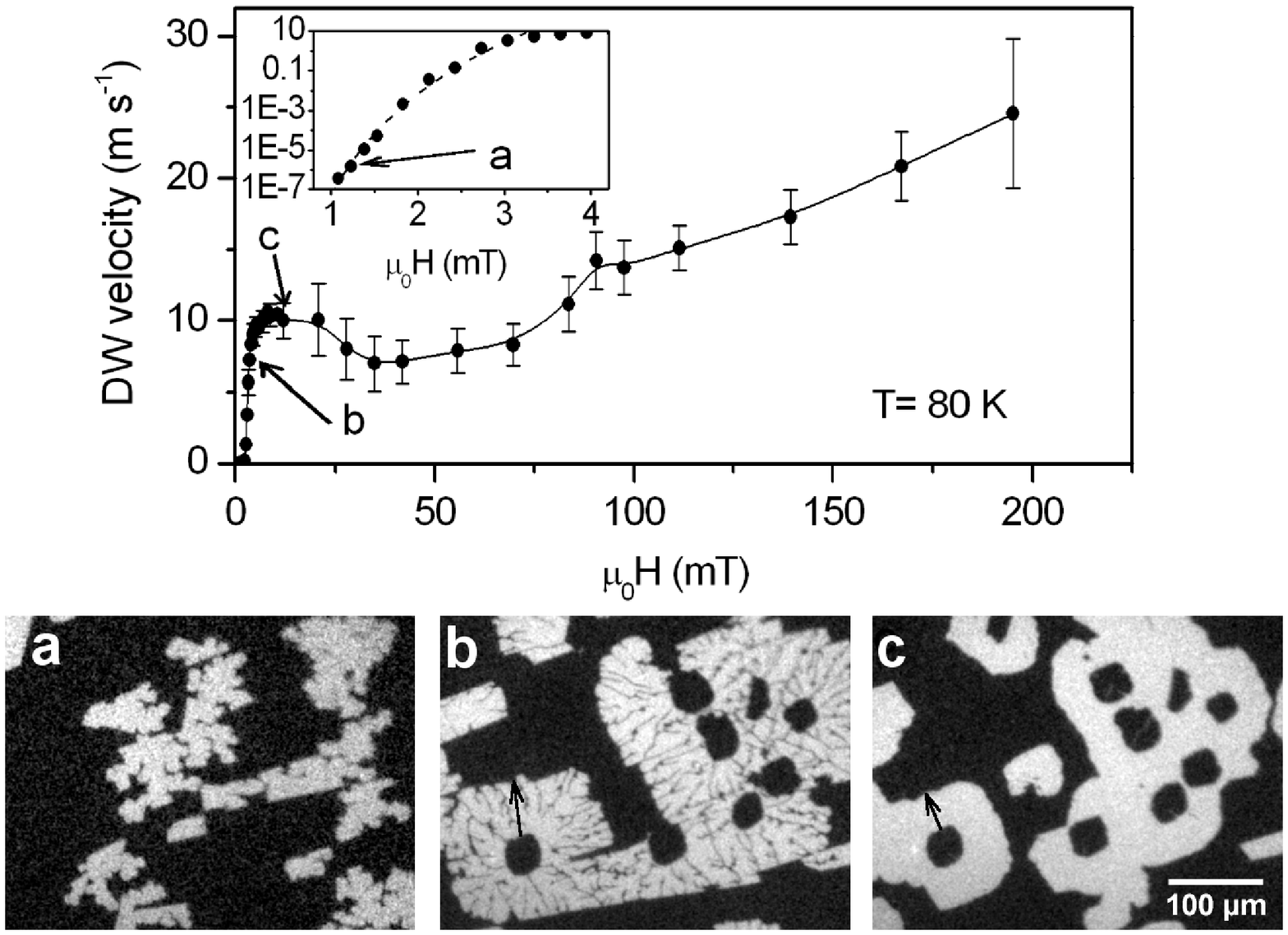}
	\end{center} 
	\caption{}
	\label{fig:Graph-vitesse-vs-champ}
\end{figure}

The sample consists of an annealed Ga$_{0.93}$Mn$_{0.07}$As epilayer of thickness $d$=50 nm grown on a relaxed Ga$_{0.902}$In$_{0.098}$As buffer deposited on a GaAs substrate. After annealing, the Curie temperature $T_C$ is 130~K and the magnetic easy axis is perpendicular to the sample plane.
Kerr microscopy is used for the direct observation of DW motion (see Ref.~\cite{dourlatJAP,thevenard} for more details). The sample is placed in a helium-flow cryostat.
The DW velocity is measured using a magnetic pulse field technique~\cite{metaxas,dourlat-spintec}.  For low velocity ($v<10^{-3}$~m~s$^{-1}$), pulses are generated by a conventional external coil (rise time $\approx 100$~ms, maximum amplitude $60$~mT). The DW velocity is determined from a set of snapshots tracking the DW position during a field plateau (see image (a) in Fig.~\ref{fig:Graph-vitesse-vs-champ}). For higher velocity ($v>10^{-3}$~m~s$^{-1}$), pulses are generated by a small coil (rise time $\approx 200$~ns, maximum amplitude $250$~mT) of diameter $\approx1$~mm placed inside the cryostat, onto the sample surface. DW motion is driven by a series of pulses of constant amplitude $H$ and increasing duration $\tau$. The snapshots recorded before and after each pulse are used to determine the DW displacement as a function of $\tau$ as shown in images (b) and (c) of Fig.~\ref{fig:Graph-vitesse-vs-champ}. The DW velocity reported in Fig.~\ref{fig:Graph-vitesse-vs-champ} corresponds to a linear fit of this curve (not shown). This procedure eliminates the effects of the pulse rise and decay times.

A typical velocity curve for field-driven DW dynamics is shown in Fig.~\ref{fig:Graph-vitesse-vs-champ} ($T=80$~K). Several dynamical regimes can be identified. 
For $1$~mT$<\mu_0H<2.5$~mT {\it - creep regime -} DWs are rough and present spatially inhomogeneous displacements (image (a) in Fig.~\ref{fig:Graph-vitesse-vs-champ}). The free DW motion is impeded by defects, some of them appearing as pointlike or long linear defects~\cite{dourlatJAP,thevenard}. To determine the average velocity, DW displacements were only measured in areas with no strong pinning defects. The results are reported in the inset of Fig.~\ref{fig:Graph-vitesse-vs-champ}. $v$ is found to increase over 6 orders of
magnitude. The good agreement with a fit $v=v_0exp(-a H^{-1/4})$ suggests that DWs follow a creep regime described by the motion of an elastic string in the presence of a random pinning potential~\cite{lemerle}. However, the systematic investigation of this regime is out of the scope of this paper.
For 2.5~mT$<\mu_0 H <8$~mT {\it - depinning regime -} the velocity varies linearly with the field. The DW roughness decreases (image (b) in Fig.~\ref{fig:Graph-vitesse-vs-champ}) and the domains expand almost isotropically. One still notices pointlike defects around which the DWs fold up, creating lamellar domains~\cite{dourlatJAP,thevenard}.
This
regime, separating the creep and flow regimes, corresponds
to the depinning regime. 
For $\mu_0 H >8$~mT {\it - flow regimes -} the DWs are smooth and the displacements homogeneous (image (c) in Fig.~\ref{fig:Graph-vitesse-vs-champ}), thereby indicating that the DW dynamics is no more limited by pinning. The $v(H)$ curve (Fig.~\ref{fig:Graph-vitesse-vs-champ}) presents the main features of the predicted dissipation-limited DW dynamics~\cite{slonczewski73-74,schryer-walker,hubert-german,malozemoff}. A velocity peak ($v=10.5$~m~s$^{-1}$ for $\mu_0 H=8.2$~mT) is followed by a region with negative differential mobility (8.2~mT$<\mu_0H<35$~mT). The DW velocity decreases down to 7~m~s$^{-1}$ for $\mu_0 H=35$~mT.
For $\mu_0H>50$~mT, the velocity increases again with the field. $v$ is proportional to $H$ as expected for the high field precessional regime, except near $\mu_0 H=92$~mT where an unpredicted second velocity peak is observed. In order to determine whether those features of the DW dynamics are systematically observed, measurements were performed over a wide temperature range (0.03$< T/T_C<$0.92).



\begin{figure}[htbp]
	\begin{center}
		\includegraphics[width=0.9\linewidth]{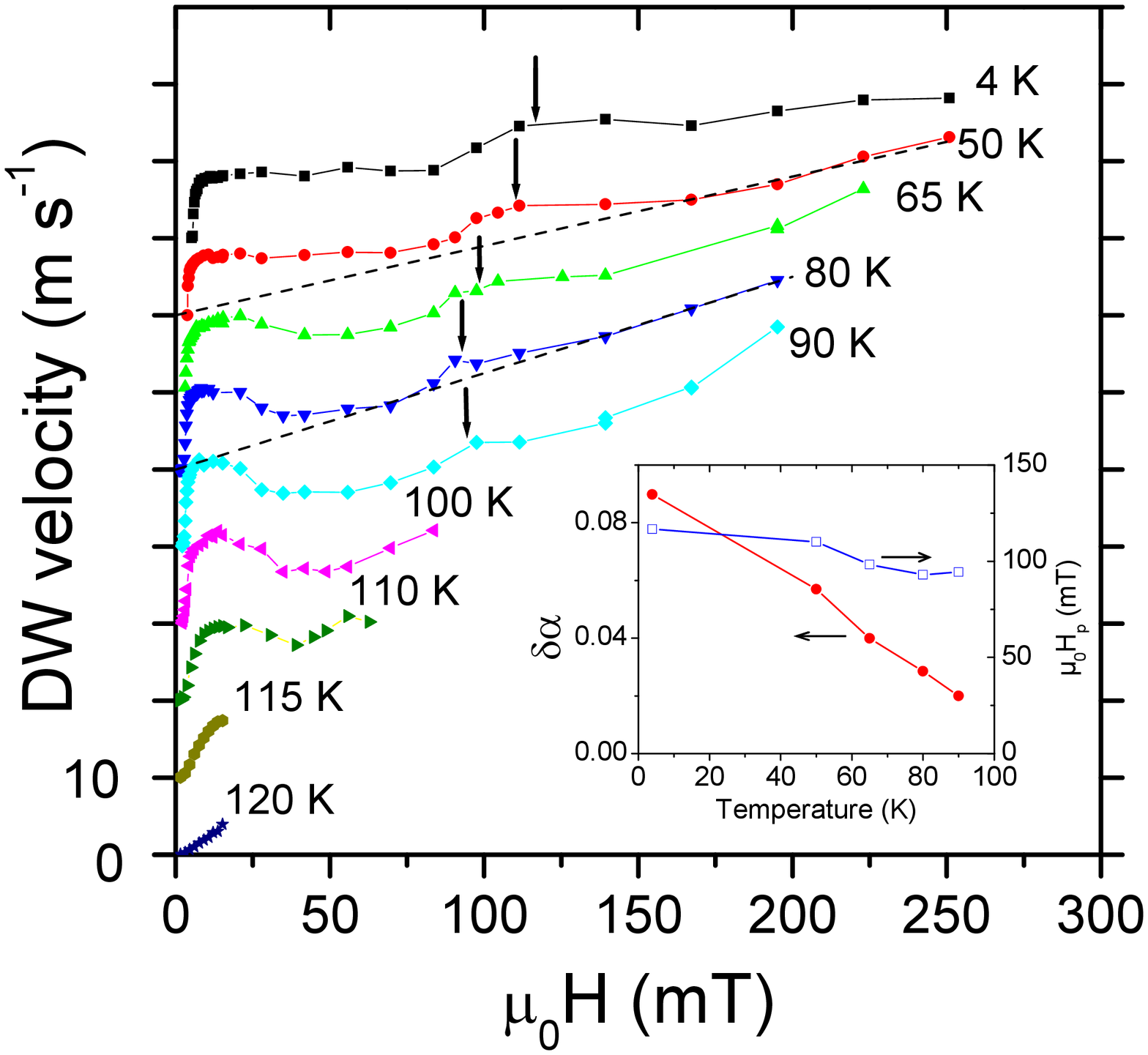}
	\end{center}
	\caption{}
	\label{fig:graph-vitesse-fct-ture}
\end{figure}

The results are reported in Fig.~\ref{fig:graph-vitesse-fct-ture}. Qualitatively, the $v(H)$ curves show a weak dependence on temperature for the depinning and flow regimes. The main difference concerns the upper boundary $H_{up}$ of the investigated field range. At low temperature ($T=4$-$50$~K), $H_{up}$ corresponds to the maximum available field amplitude (250~mT). At higher temperature ($T=65$-$120$~K) $H_{up}$ is limited by nucleation. The distance between domains nucleated during a pulse is too small for an accurate measurement of the DW displacement. The anomalous velocity peak in the high field precessional regime is systematically observed, as shown by the arrows in Fig.~\ref{fig:graph-vitesse-fct-ture}.

In the following the main features of the DW dynamics are analyzed quantitatively. In uniaxial ferromagnetic films, the anisotropy is characterized by the quality factor $Q = 2 K_u/\mu_0 M_s^2$, where $K_u$ is the uniaxial anisotropy constant and $M_s$ the saturation magnetization~\cite{malozemoff}. Q is found in the range 8.6-14, which denotes strong uniaxial anisotropy~\cite{gourdonPRBmicromagnetic}. In that case, the velocity in the Walker steady regime is given by $v_{st}(H)= \mu_{st}\mu_0 H = \gamma \Delta \mu_0 H/\alpha$, where $\gamma$ is the gyromagnetic factor (1.76~10$^{11}$~Hz~T$^{-1}$), $\Delta$ the DW width parameter, and $\alpha$ the damping parameter~\cite{malozemoff}. In the high field range of the precessional regime the velocity is given by $v_{prec}(H)=\mu_{prec}\mu_0 H =\gamma \Delta \mu_0 H \alpha /(1+\alpha^2)$~\cite{malozemoff}. 

\begin{figure}[htbp]
	\begin{center}
		\includegraphics[width=0.95\linewidth]{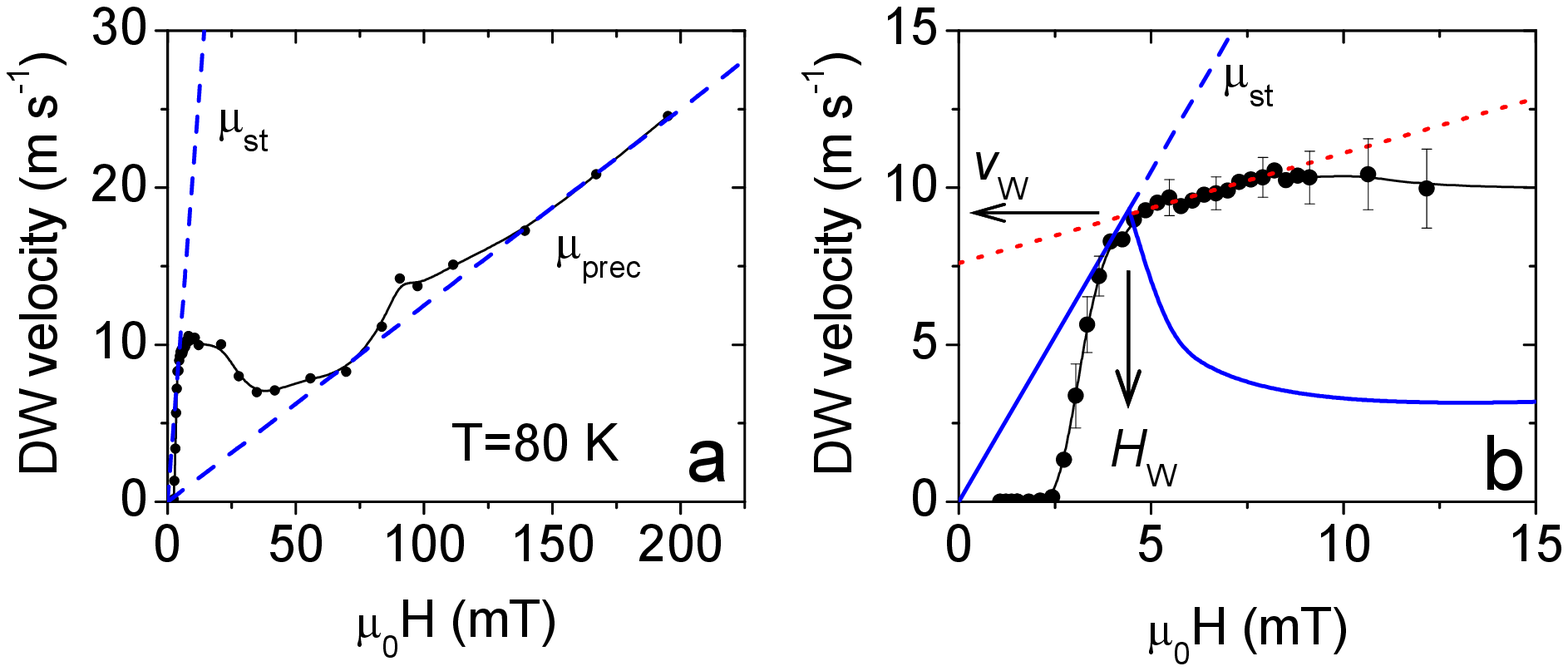}
	\end{center}
	\caption{}
	\label{fig:fit-80K}
\end{figure}

\begin{figure}[htbp]
	\begin{center}
		\includegraphics[width=0.95\linewidth]{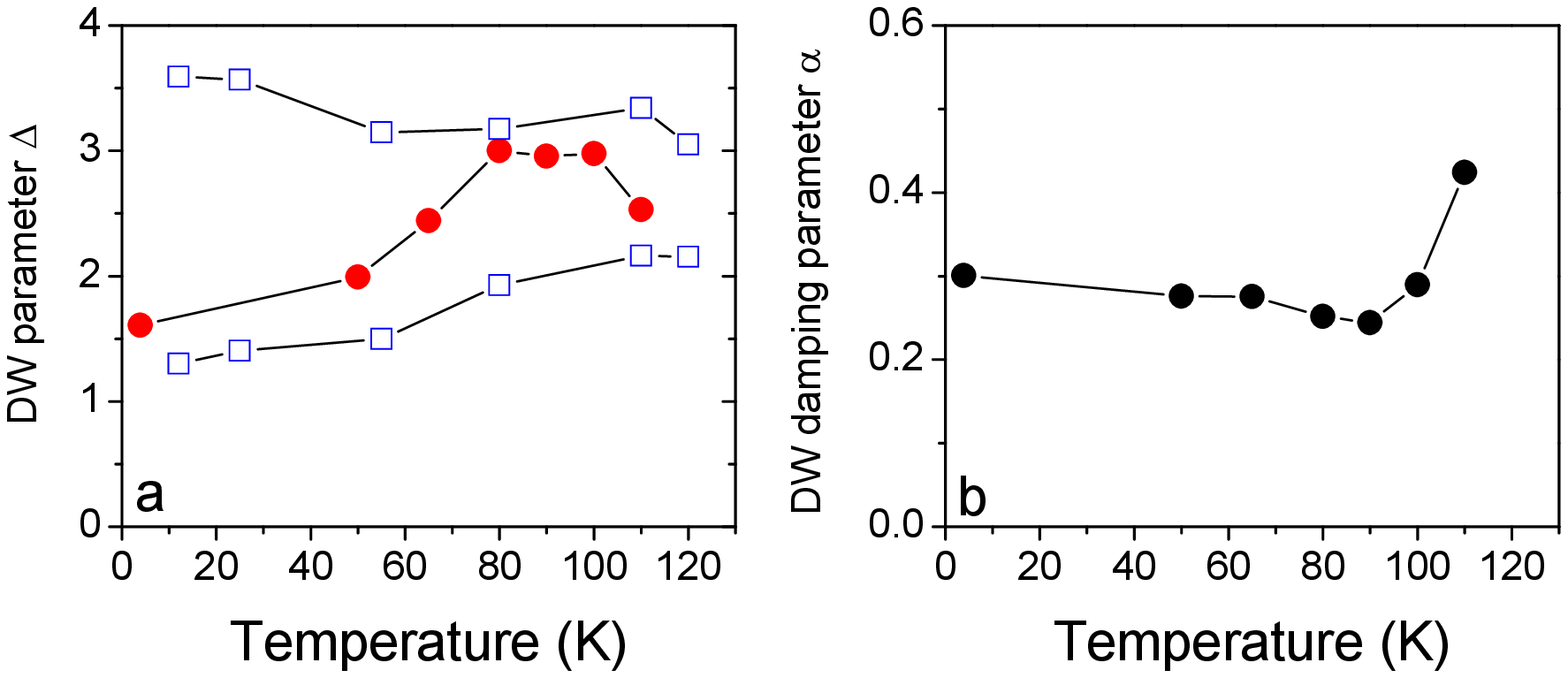}
	\end{center}
	\caption{}
	\label{fig:Delta-Alpha}
	\end{figure}

Those predictions are compared to the observed dynamical regimes. Since the flow regime is reached close to the velocity peak the mobility $\mu_{st}$ is obtained by adjusting a straight line tangent to the experimental curve (Fig.~\ref{fig:fit-80K} (a) and (b)). $\mu_{prec}$ is obtained from a linear fit of the high-field regime (Fig.~\ref{fig:fit-80K}(a)).  The ratio $\mu_{prec}/\mu_{st}=\alpha^2/(1+\alpha^2)$ yields $\alpha$. $\Delta$ is obtained as $\Delta=\mu_{prec}(1+\alpha^2)/\alpha \gamma$. As shown in Fig.~\ref{fig:Delta-Alpha}(a)
$\Delta$ varies weakly with the temperature. Moreover, the $\Delta(T)$ values lie between the two boundaries deduced independently from domain theory for the same sample~\cite{gourdonPRBmicromagnetic}. This very good quantitative agreement confirms that the procedure used for the determination of $\Delta$ and $\alpha$ is relevant. This also demonstrates that the precessional and steady regimes are indeed observed experimentally, the latter being reached only near the Walker velocity peak as shown in Fig.~\ref{fig:fit-80K}(b). 

The damping coefficient $\alpha \approx 0.3$ is also weakly dependent on the temperature (Fig.~\ref{fig:Delta-Alpha}(b)). A similar value ($\alpha \approx 0.15$) was obtained from the decay of the magnetization precession induced by optical excitation~\cite{Qi}. Surprisingly, the DW damping is more than one order of magnitude larger than the damping deduced from the frequency dependence of the ferromagnetic resonance (FMR) linewidth: $\alpha\approx 0.01$ for this sample~\cite{khazen-tobepublished}, a value in agreement with theoretical predictions for GaMnAs ($\alpha\approx 0.02-0.03$)~\cite{sinova}. A similar discrepancy was reported for garnet films with small $\alpha$-values~\cite{vella-coleiro,thiavilleEPJ}. 
Theoretically, the DW equation of motion is derived from the Landau-Lifshitz-Gilbert (LLG) equation for the magnetization. The dissipation is classically described by the phenomenological Gilbert damping coefficient $\alpha$ that accounts for the relaxation of the direction of the magnetization vector~\cite{slonczewski73-74,schryer-walker,hubert-german,malozemoff}. However, dissipation processes are expected to differ for uniform magnetization (FMR) and for a moving DW. A deviation from the linear steady regime is predicted when the interaction of the moving DW with thermal magnons is taken into account~\cite{ivanov}. However, this interaction can be safely discarded here since it would give a significant contribution only at high velocity, above $\approx$ 300~m~s$^{-1}$ for this sample. In the low velocity range, additional relaxation terms in the LLG equation, taking into account the non-conservation of the magnetization modulus and heat exchange with a thermostat, can lead to a DW dynamical damping larger than the Gilbert (FMR) damping~\cite{baryakhtar,huang}. For a magnetization response time $\tau_M=\chi_{\|}/\gamma M_s\alpha_{FMR}$ shorter than the DW transit time $\tau_t=\Delta/v$, the dissipation for DW motion in the steady regime is described by a dynamical damping~\cite{huang} 
	$\alpha_{DW}=\alpha_{FMR}\left[1+16 \left(4\pi\chi_{\|}Q\right)^2/3\alpha_{FMR}^2\right]$, 
where $\chi_{\|}$ is the longitudinal magnetic susceptibility (CGS units). Using this equation with $\alpha_{FMR}\approx 0.01$ yields $\chi_{\|}\approx 10^{-4}$, in reasonable agreement with theoretical estimations based on spin-wave theory~\cite{holstein,argyle}. Taking into account the anisotropy gap, one finds $\chi_{\|}$ in the range  $4~10^{-6}$-$10^{-5}$ for $T=12$~K and $8~10^{-5}$-$2~10^{-4}$ for $T=80$~K. For the measured velocities one finds $\tau_M\ll\tau_t$, thereby justifying the use of this model. We point out that the theoretical prediction of enhanced DW dynamical damping in the steady regime also accounts well for the damping in the high-field precessional regime.

Let us now extend the analysis of the experimental velocity curves beyond the 1D theory of DW motion. As calculated and shown in Fig.~\ref{fig:fit-80K}(b), the 1D theory predicts a maximum velocity (Walker velocity $v_W=\gamma\Delta\mu_0 M_s/2$) at the Walker field $H_W=\alpha M_s/2$. For $H$ just above $H_W$ the DW back and forth motion due to the DW magnetization precession around the applied field should lead to a decrease of the time-averaged velocity and hence to a region of negative differential mobility~\cite{hubert-german}. In contrast, the experimental $v(H)$ curve just shows a change of slope with still a constant positive mobility, yet strongly reduced with respect to the steady regime (Fig.~\ref{fig:fit-80K}(b)). In this field range, the DW structure is expected to be unstable. A solution for DW propagation with generation and propagation of horizontal Bloch lines through the film has been proposed~\cite{slonczewski73-74}. For sufficiently large $\alpha$ this model predicts a viscous-like drag with decreased mobility $\mu_{BL}=\gamma\Delta/\left[\alpha\left(1+\pi^2\Lambda/2\alpha^2a\right)\right]$ with $\Lambda=\Delta\sqrt{Q}$ the exchange length and $a$ of the order of the film thickness~\cite{slonczewski73-74}. The linear fit of the velocity curve just above the Walker field for the set of investigated temperatures yields $a$ in the range 86$\pm$50~nm, consistent with the sample thickness $d=50$~nm. Given the fact that the exchange length $d/10<\Lambda<d/5$ is not very much smaller than $d$, contrary to the assumption of the model, the agreement is quite satisfactory. It strongly suggests that DW motion above the Walker field is slowed down by the repetitive generation of one Bloch line at one surface of the film and subsequent propagation and annihilation at the other surface. 

Let us now discuss the intriguing velocity peak observed around 90-120 mT (Fig.~\ref{fig:graph-vitesse-fct-ture}). To our knowledge, such a peak has been neither predicted nor observed in ferromagnetic systems. Since it occurs within the linear precessional regime, where the velocity is proportional to the damping parameter, it can be ascribed to a damping resonance. In order to characterize its temperature dependence, the velocity curves are fitted using a field-dependent damping $\alpha(H)=\alpha+\delta\alpha \exp\left[-\left(H-H_p\right)^2/2\sigma^2\right]$. As shown in Fig.~\ref{fig:graph-vitesse-fct-ture} (inset) the resonance amplitude ($\delta\alpha$=0.04 at 60 K) decreases linearly with the temperature, extrapolating to zero at $T\approx120$~K, close to $T_C$. The resonance field $H_p$ varies slightly with the temperature. The corresponding energy $\hbar\omega=\hbar\gamma \mu_0\sqrt{H^2-H_W^2}/(1+\alpha^2)$ is of the order of 10~$\mu$eV ($\omega=17$~GHz). It may correspond to transitions between confined states of volume magnons, as calculated from the dispersion curve using the spin stiffness constant determined from this work and ref.~\cite{gourdonPRBmicromagnetic}. The damping resonance might also be related to DW excitations (flexural modes), whose energy spectrum lies inside the anisotropy gap of the volume magnons (43~$\mu$eV at 80~K). These excitations have been calculated for a static or moving DW in the steady regime~\cite{schlomann,thiele} but not in the precessional one.

Despite the small value of the saturation magnetization and hence of the Walker field in GaMnAs, we could observe the dissipation-limited flow regimes beyond the creep and depinning regimes. The steady as well as the precessional regime are described by a DW dynamical damping, which is found to be much larger than the FMR damping, in agreement with theoretical predictions considering non-conservation of the magnetization modulus. However, contrary to the classical assumption, a single, field-independent, damping constant cannot account for the whole DW dynamics. The existence of a damping resonance inside the precessional regime suggests an additional dissipation mechanism and calls for further theoretical investigations of dissipation processes in this regime.


\begin{acknowledgments}
We gratefully acknowledge fruitful discussions with Jacques Ferr\'{e}, Alain Mauger and Andrejs Cebers. This work was in parts supported by R$\acute{e}$gion Ile de France under contract IF07-800/R with C'Nano IdF at CNRS.
\end{acknowledgments}

\newpage
\begin{center}
FIGURE CAPTIONS
\end{center}
FIG. 1 Magnetic field dependence of the DW velocity at $T=80$~K. The error bars represent the standard deviation of the velocity distribution. Inset: low field region in semi-logarithmic scale with a fit (dashed line) according to the creep model~\cite{lemerle}. Labels (a), (b), and (c) refer to the corresponding images in the creep, depinning and flow regimes, respectively. Images (b) and (c) are differential images showing the DW displacement as indicated by black arrows.\\

FIG. 2 (Color online) Temperature dependence of the DW field-velocity curves. Each curve is up-shifted by 10~m~s$^{-1}$ with respect to the previous one. Inset: temperature dependence of the resonance field $H_p$ and damping $\delta\alpha$ of the velocity peak indicated by arrows.\\


FIG. 3 (Color online) Comparison between theoretical predictions and experimental results (black circles). Dashed blue lines in (a): linear velocity in the steady and precessional regimes with mobility $\mu_{st}$ and $\mu_{prec}$, respectively. Solid blue curves in (b): velocity in the steady regime (below $H_W$) and in the unstable precessional regime (above $H_W$) calculated using the experimentally determined values of $\Delta$ and $\alpha$ (this work) and $M_s$~\cite{gourdonPRBmicromagnetic}; dotted red line: linear fit of the velocity in the Bloch line regime.\\


FIG. 4 (Color online) (a) Comparison between the DW parameter $\Delta$ obtained from DW dynamics (full circles) and the boundaries for $\Delta$ obtained from domain theory (open symbols)~\cite{gourdonPRBmicromagnetic}. (b) Temperature dependence of the DW damping parameter $\alpha$.

\newpage

\newpage

\newpage

\newpage
	
\end{document}